\documentclass[letterpaper,english,reprint,superscriptaddress]{revtex4-1}
\usepackage[T1]{fontenc}
\usepackage[utf8]{inputenc}
\usepackage{babel}
\usepackage{textcomp}
\usepackage{amsmath}
\usepackage{amssymb}
\usepackage{graphicx}
\usepackage{wasysym}
\usepackage[unicode=true,pdfusetitle,
 bookmarks=false,
 breaklinks=false,pdfborder={0 0 0},backref=false,colorlinks=false]
 {hyperref}

\makeatletter

\pdfpageheight\paperheight
\pdfpagewidth\paperwidth

\makeatother

\begin{document}

\title{Purification of a single photon nonlinearity}

\author{H. Snijders}

\affiliation{Huygens--Kamerlingh Onnes Laboratory, Leiden University, P.O. Box
9504, 2300 RA Leiden, The Netherlands}

\author{J. A. Frey}

\affiliation{Department of Physics, University of California, Santa Barbara, California
93106, USA}

\author{J. Norman}

\affiliation{Department of Electrical \& Computer Engineering, University of California,
Santa Barbara, California 93106, USA}

\author{M. P. Bakker}

\affiliation{Huygens--Kamerlingh Onnes Laboratory, Leiden University, P.O. Box
9504, 2300 RA Leiden, The Netherlands}

\author{A. Gossard}

\affiliation{Department of Electrical \& Computer Engineering, University of California,
Santa Barbara, California 93106, USA}

\author{J. E. Bowers}

\affiliation{Department of Electrical \& Computer Engineering, University of California,
Santa Barbara, California 93106, USA}

\author{M. P. van Exter}

\affiliation{Huygens--Kamerlingh Onnes Laboratory, Leiden University, P.O. Box
9504, 2300 RA Leiden, The Netherlands}

\author{D. Bouwmeester}

\affiliation{Huygens--Kamerlingh Onnes Laboratory, Leiden University, P.O. Box
9504, 2300 RA Leiden, The Netherlands}

\affiliation{Department of Physics, University of California, Santa Barbara, California
93106, USA}

\author{W. Löffler}

\affiliation{Huygens--Kamerlingh Onnes Laboratory, Leiden University, P.O. Box
9504, 2300 RA Leiden, The Netherlands}
\begin{abstract}
We show that the lifetime-reduced fidelity of a semiconductor quantum
dot--cavity single photon nonlinearity can be restored by polarization
pre- and postselection. This is realized with a polarization degenerate
microcavity in the weak coupling regime, where an output polarizer
enables quantum interference of the two orthogonally polarized transmission
amplitudes. This allows us to transform incident coherent light into
a stream of strongly correlated photons with a second-order correlation
function of $g^{2}(0)\apprge40$, larger than previous experimental
results even in the strong-coupling regime. This purification technique
might also be useful to improve the fidelity of quantum dot based
logic gates.

\end{abstract}
\maketitle
\textbf{}

Single photon nonlinearities enabled by quantum two-level systems
are essential for future quantum information technologies, as they
are the building block of quantum photonics logic gates \cite{imamoglu1997},
deterministic entanglers of independent photons \cite{Bonato2010},
and for coupling distant nodes to form a quantum network \cite{Kimble2008}.
Near unity fidelity interaction of photons with a two level system
such as an atom or quantum dot (QD) is enabled by embedding it into
an optical cavity \cite{Imamoglu1999}, where the electronic and photonic
states become bound and form the dressed states \cite{jaynes1963}
of cavity quantum electrodynamics (CQED). A hallmark of single photon
nonlinearities is the modification of the photon statistics of a quasi-resonant
weak coherent input beam \cite{kubanek2008}: First, the transmitted
light photon statistics can become antibunched due to the photon blockade
effect \cite{imamoglu1997,birnbaum2005,dayan2008}, which is enabled
by the anharmonicity of the Jaynes-Cummings ladder \cite{schuster2008,kasprzak2010,faraon2010b}.
Second, the system can be tuned to reach the regime of photon tunnelling
\cite{faraon2008,kubanek2008} where the single-photon component is
reduced and photons are transmitted in $N>1$ Fock states or ``photon
bundles'' \cite{valle2012,munoz2014}.

\begin{figure}
\includegraphics[width=1\columnwidth]{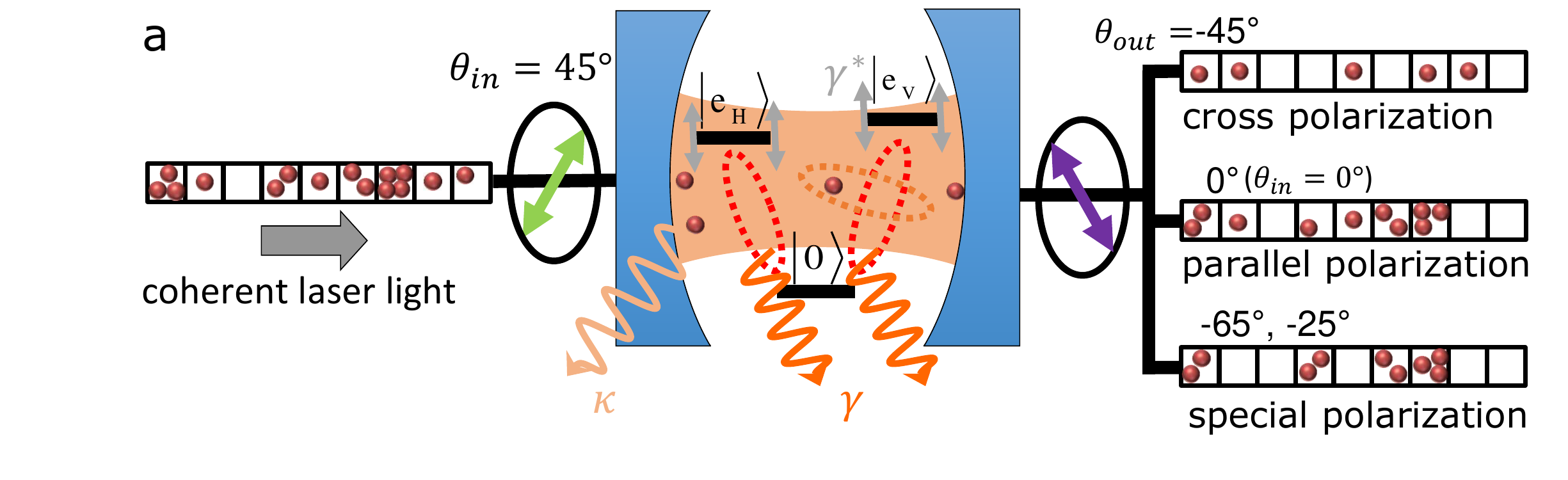}

\includegraphics[width=0.6\columnwidth,height=4cm]{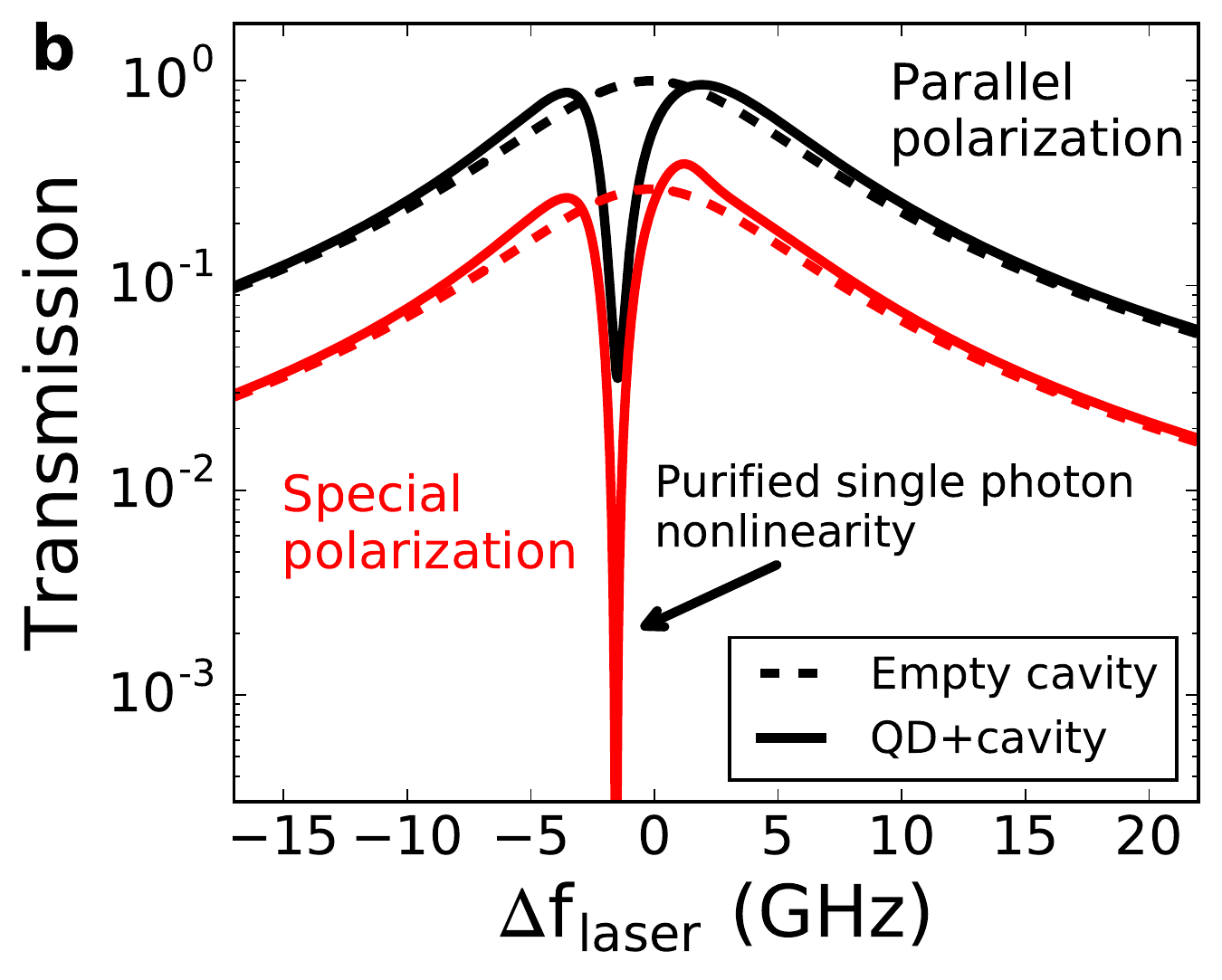}\includegraphics[width=0.4\columnwidth,height=4.1cm]{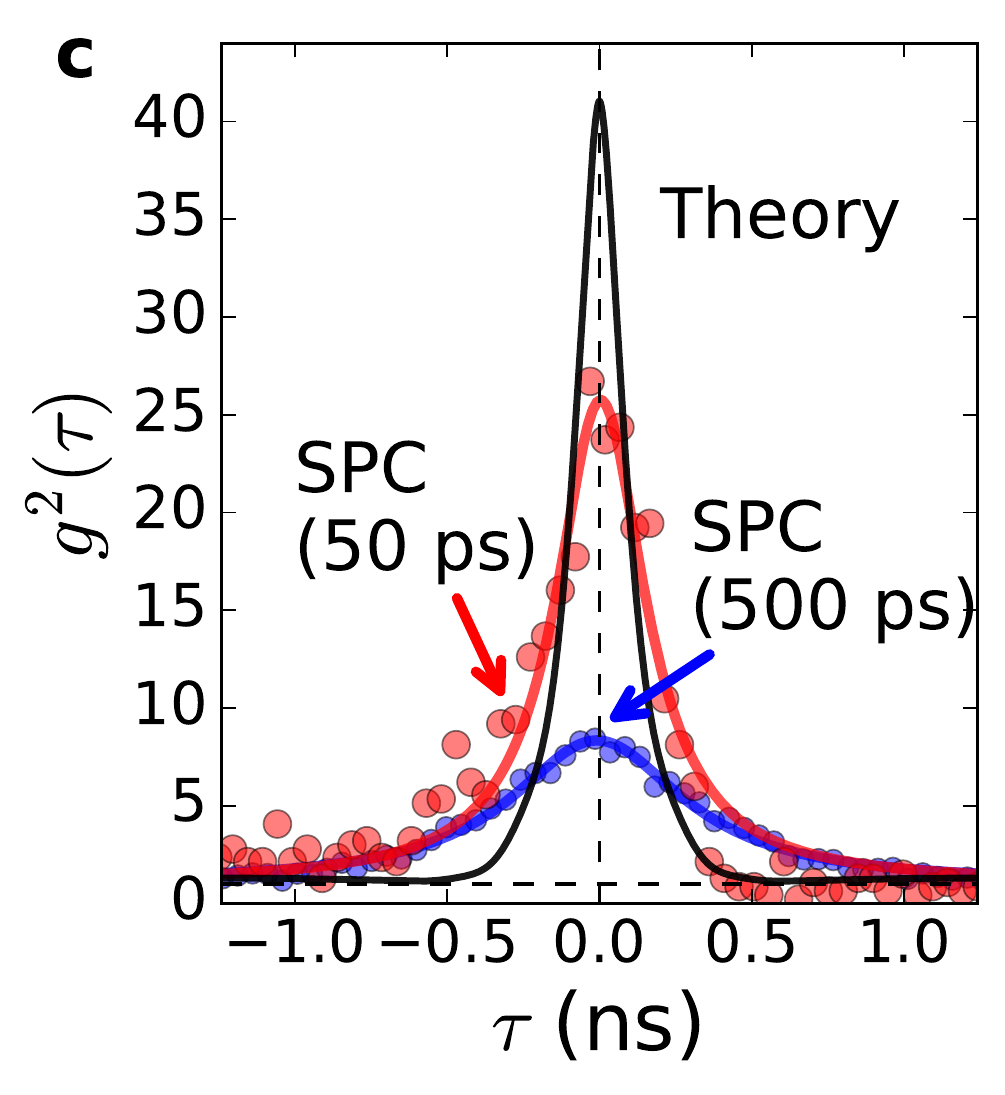}

\caption{\label{fig:introfig} (a) Cartoon of the experiment: Polarization
pre- and postselection in a resonant transmission CQED experiment
enables tuning of the photon statistics from antibunched to bunched.
(b) Theoretical resonant transmission spectra for coherent light with
mean photon number $\ll1$, with and without the quantum dot, comparing
the conventional case (parallel polarizers) to the case of special
polarization postselection along $\theta_{out}^{\ast}$: close to
one of the QD resonances, single-photon transmission is perfectly
suppressed, despite the finite lifetime and cavity coupling of the
QD transition. (c) Second-order correlation function (QD B) for the
special polarization angle case, comparing theory and experiment using
two different sets of single photon counters (SPCs) with different
timing jitter, 50~ps and 500~ps.}
\end{figure}

In terms of the second-order photon correlation function $g^{2}(0)$,
values up to $\sim2$ \cite{reinhard2012,majumdar2012,rundquist2014,mueller2015}
have been obtained experimentally with quantum dots, which hardly
exceeds even the classical case of thermal light following Bose statistics
of $g^{2}(0)=2$. In atomic systems with much longer coherence times,
values up to $\sim50$ have been obtained \cite{kubanek2008}, and
it is known \cite{mcneil1975} that strict two-photon light sources
exhibit diverging $g^{2}(0)$ if the two-photon flux is reduced. Most
related quantum dot experiments to date have been operating in the
strong-coupling regime of CQED, which is considered to be essential
due to its photon-number dependent energy structure \cite{rundquist2014,mueller2015,kubanek2008}.
However, strong coupling requires a small optical mode volume, which
in turn makes it extremely hard to achieve polarization degeneracy
of the fundamental cavity mode. This is due to unavoidable deviations
from the ideal shape and intrinsic birefringence \cite{Yoshie2004,Reithmaier2004}
on the GaAs platform, precluding implementation of deterministic polarization-based
quantum gates \cite{hu2009,Bonato2010,arnold2015}. Here we show,
using a polarization degenerate cavity in the weak coupling CQED regime,
that we can transform incident coherent light into a stream of strongly
correlated photons with $g^{2}(0)\apprge25$, corresponding to $\gtrsim40$
in the absence of detector jitter. The polarization degenerate cavity
enables us to choose the incident polarization $\theta_{in}=45^{\circ}$
such that both fine-structure split quantum dot transitions along
$\theta_{QD}^{X}=90^{\circ}$ and $\theta_{QD}^{Y}=0^{\circ}$ are
excited, and we can use a postselection polarizer behind the cavity
($\theta_{out}$) to induce quantum interference of the two transmitted
orthogonal polarization components (Fig.~\ref{fig:introfig}a). This
leads to the appearance of two special postselection polarizer angles
$\theta_{out}^{\ast\pm}$ (depending on sample parameters), which
can be used to restore perfect QD contrast (one of them is shown in
Fig.~\ref{fig:introfig}b). This compensates fully for reduced QD-cavity
coupling due to finite QD lifetime and QD-cavity coupling strength,
leading to complete suppression of transmission of the single-photon
component in the low excitation limit. The transmission of higher-photon
number states remains largely intact, allowing us to observe in Fig.~1c
the strongest photon correlations to date in a quantum dot system,
even exceeding those of strongly coupled atomic systems \cite{kubanek2008}.
In the following a detailed experimental and theoretical investigation
of this effect, which can be seen as a purification of a single-photon
nonlinearity, will be presented.

\textbf{}

\textbf{}

\section*{\label{results}Results}

\noindent \textbf{Device structure. }Our device \footnote{See Supplemental Material.}
consists of self-assembled InAs/GaAs quantum dots embedded in a micropillar
Fabry-Perot cavity grown by molecular beam epitaxy \cite{Strauf2007}.
The QD layer is embedded in a P-I-N junction, which enables tuning
of the QD resonance frequency by the quantum confined Stark effect.
For transverse mode confinement and to achieve polarization degenerate
cavity modes, we first ion-etch micropillars of large diameter (35~\textmu m)
and slightly elliptical shape, then we use wet-chemical oxidation
of an AlAs layer \cite{Bakker2014} to prepare an intra-cavity lens
for transverse-mode confinement \cite{coldren1996}, avoiding loss
by surface scattering at the side walls. Finally, we fine-tune the
cavity modes by laser induced surface defects \cite{Bonato2009,bakker2015}
to obtain routinely a polarization mode splitting much less than 10\%
of the cavity linewidth. We show here data from two quantum dots,
QD A and B, where QD A (B) is separated by a 20~nm (35~nm) tunnel
barrier from the electron reservoir.

\medskip{}

\noindent \textbf{Device parameters and theoretical model. }The system
we study here is tuned to contain a single neutral QD within the cavity
linewidth. The excitonic fine structure splitting leads to $\approx2.5$~GHz
splitting between the orthogonally polarized QD transitions at 0$^{\circ}$
($\omega_{QD}^{Y}$) and 90$^{\circ}$ ($\omega_{QD}^{X}$). To determine
further system parameters, we model our QD-cavity system by a two-polarization
Jaynes-Cummings (JC) Hamiltonian coupled to the incident coherent
field, and take care of cavity and QD dissipation by the quantum master
equation formalism \cite{armen2006,Johansson2013}. We compare experiment
and theory for 6 different input-output polarizer settings to faithfully
determine the model parameters \cite{Note1}. For QD A we obtain a
cavity decay rate $\kappa=69$~ns$^{-1}$, QD relaxation rate $\gamma_{||}=3.5$~ns$^{-1}$,
QD pure dephasing $\gamma^{*}=6$~ns$^{-1}$ , and QD-cavity coupling
rate $g=15$~ns$^{-1}$. These measurements were performed for an
input power of 100~pW to avoid saturation effects. With this we can
calculate the device cooperativity $C=\frac{g^{2}}{\kappa\gamma}=0.4$
(with $\gamma=\frac{\gamma_{||}}{2}+\gamma^{*}$), which puts our
system in the weak coupling regime. For QD B {[}Fig.~1 (c){]} we
obtain $\kappa=105$ $ns^{-1}$, $g=14$ $ns^{-1}$, $\gamma^{||}\thickapprox1.0$
$ns^{-1}$ $\pm$ $0.3$ $ns^{-1}$, $\gamma^{*}\thickapprox0.7$
$ns^{-1}$$\pm$ $0.3$ $ns^{-1}$, leading to a enhanced cooperativity
$C\approx1.6\pm0.5$ ns$^{-1}$. Details are given in the supplemental
information \cite{Note1}. The cavity with QD B shows a residual polarization
splitting of 4~GHz, which leads to less-perfect agreement of our
theoretical model. Therefore we focus now on a different cavity with
QD A having a lower maximal $g^{2}(0)\approx3.7$ but with excellent
agreement to theory.

\medskip{}

\noindent \textbf{Resonant photon correlation spectroscopy. }We use
a narrowband (100~kHz) laser to probe the system and study the transmitted
light (Fig.~\ref{fig:introfig}a), as a function of incident polarization,
frequency, and postselection polarizer angle behind the cavity. For
each set of parameters, we measure the resonantly transmitted light
intensity and its second-order photon correlation function $g^{2}(\tau)$
using a Hanbury Brown Twiss setup. The discrete nature of the QD levels
leads to a strongly nonlinear response of the system depending on
the incident photon number distribution; we operate at low intensities
to avoid saturation effects. We show here only data for an incident
polarization $\theta_{in}=45^{\circ}$, under which angle both QD
transitions are equally excited, additional data is given in the supplemental
information \cite{Note1}.

\noindent 

\begin{figure}
\includegraphics[width=1\columnwidth]{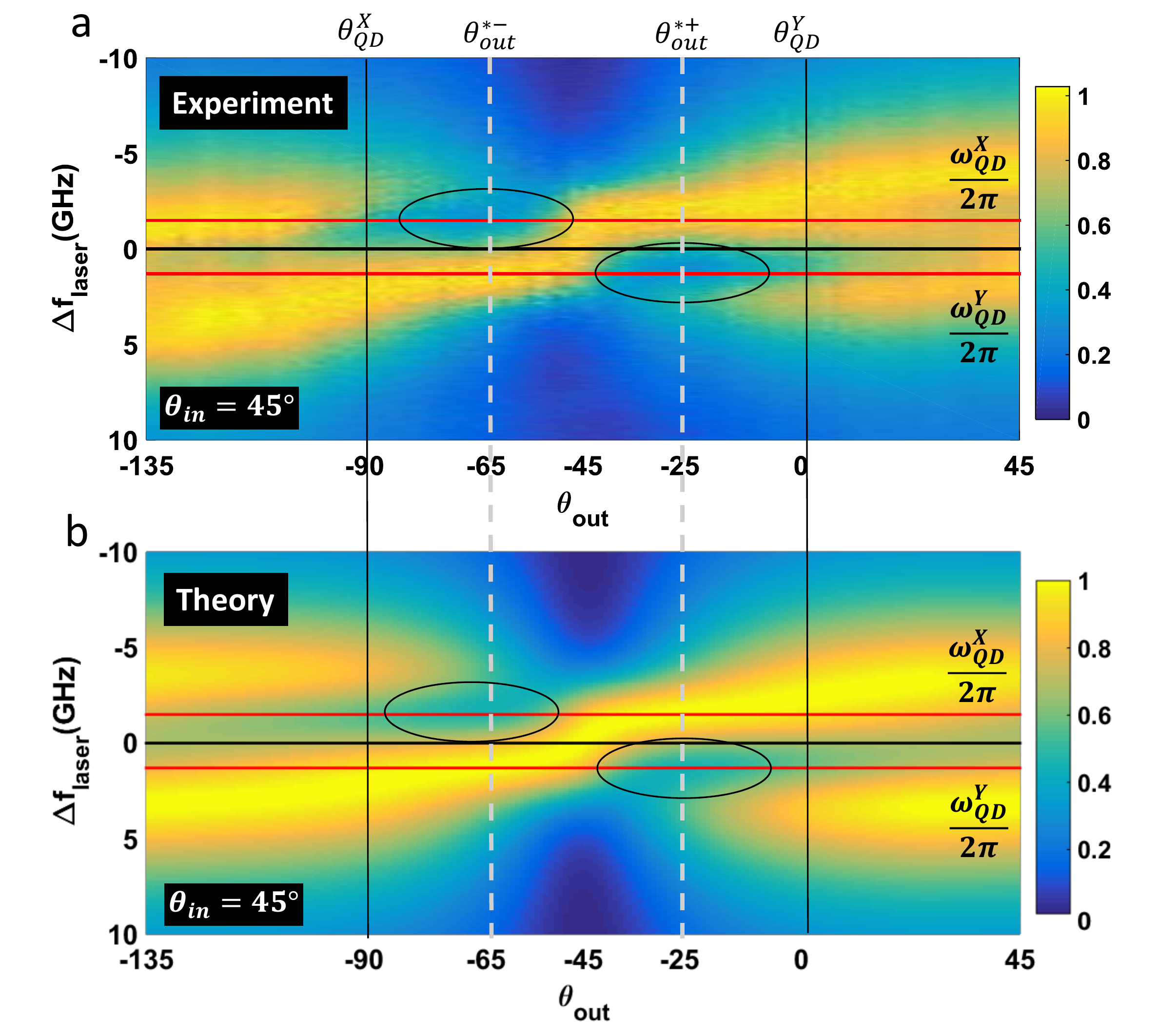}

\caption{\label{fig:xpolmagic2d}Experimental (a) and theoretical (b) false
color plot of the columnwise normalized optical transmission as a
function of the laser detuning $\Delta f_{Laser}$ and the polarization
$\theta_{out}$ ($\theta_{in}=45^{\circ}$). The fine-split QD transition
frequencies are at $f_{QD}^{X}=$$-1.5$ GHz, $f_{QD}^{Y}=$ $1.3$
GHz. The black circles indicate the special polarization conditions
with lowest transmission. }
\end{figure}

\noindent First, we compare experimental and theoretical resonant
transmission measurements in Fig.~\ref{fig:xpolmagic2d}, where the
coherent-light transmittivity as a function of the laser detuning
and orientation of the output polarizer angle $\theta_{out}$ is shown.
For clarity, we have normalized the traces for each polarization setting.
The horizontal red lines indicate the QD fine structure split transitions
$(\omega_{QD}^{X},$ $\omega_{QD}^{Y})$, the black circles indicate
regions of low transmission and the vertical dashed lines the special
polarization angles $\theta_{out}^{\ast+}\approx$$-25^{\circ}$,
$\theta_{out}^{\ast-}\approx$$-65^{\circ}$. From comparision of
both panels in Fig. \ref{fig:xpolmagic2d},\textbf{ }we find excellent
agreement between experiment and theory.

\begin{figure}
\includegraphics[width=1\columnwidth]{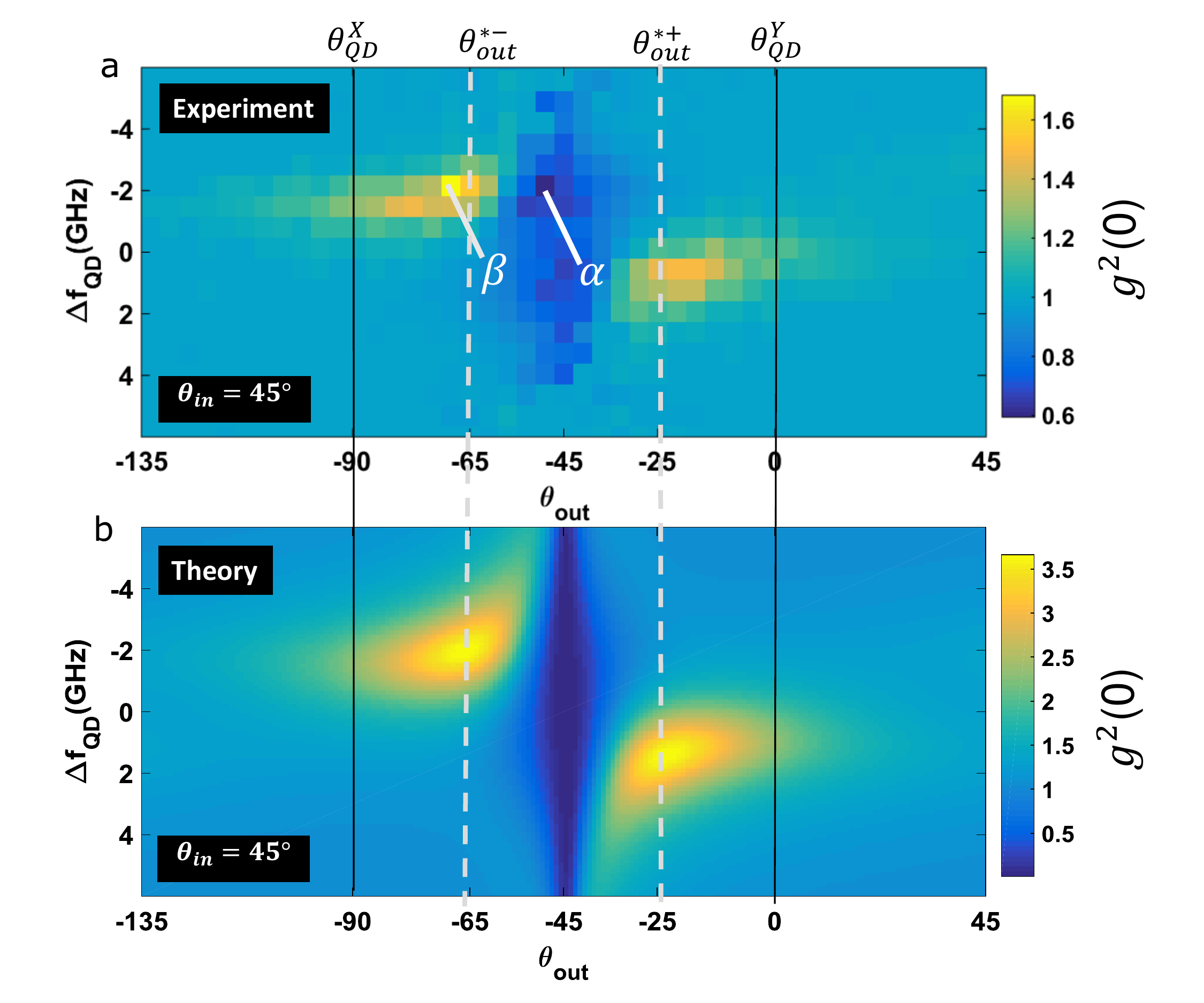}

\caption{\label{fig:g2-2d}Experimental (a) and theoretical (b) data of the
second-order correlation function as a function of the QD frequency
and output polarization ($\theta_{out}$). Green dashed lines indicate
the special polarization angles, and $\alpha$ and $\beta$ indicate
the highest and lowest experimental $g^{2}(0)$ values, respectively. }
\end{figure}

Now we perform photon correlation measurements; instead of tuning
the laser, we now tune the QD. Because the cavity linewidth is large
compared to the QD tuning range, a parameter space similar to that
before is explored. Experimentally, using an external electric field
to tune the QD via the quantum confined Stark effect is much more
robust than laser frequency tuning. Fig. \ref{fig:g2-2d} shows the
false-color map of $g^{2}(0)$ as function of output polarization
$\theta_{out}$ and QD detuning. We see clearly that the enhanced
bunching occurs under the special polarization condition in the low-transmittivity
regions indicated in Fig.~\ref{fig:xpolmagic2d}. This is expected
as in weak coherent light beams, the $P_{1}$ single-photon component
is dominating, and removal thereof should lead to enhanced bunching.
The theoretical simulation (Fig. \ref{fig:g2-2d}b) shows a maximal
photon bunching of $g^{2}(0)\approx3.7$. Compared to this, the experimentally
observed photon correlations are less (maximally $g^{2}(0)\approx2.1$),
but if we convolute the theoretical results with the detector response
function, very good agreement is obtained (see supplemental information
\cite{Note1}). One also notices that the regions of high photon bunching
exhibit hyperbolic shape, which is due to modification of the interference
between the cavity and QD resonances while scanning the QD. 

\medskip{}

\section*{discussion}

We have shown by experiment and theory that the reduced fidelity of
a QD nonlinearity, caused by imperfect QD-cavity coupling, can be
strongly enhanced by pre- and post-selection of specific polarization
states. This enables transformation of a weak coherent input beam
into highly bunched light with $g^{2}(0)\gtrsim40$, a value that
has not been reached before, not even in the strong coupling regime.
To gain more insight, understanding in terms of the photon number
distribution $P_{n}$ is required, for which we use our theoretical
model as direct experimental determination of $P_{n}$ is strongly
complicated by its sensitivity to loss. But also the simulation of
narrow-band photon number Fock input states is challenging in the
quantum master model \cite{giesz2015b}. We continue here using coherent
input light, and analyze the intra-cavity light in terms of its photon
number distribution, however, taking full care of quantum interference
at the postselection polarizer. This polarizer leads to Hong-Ou-mandel
type photon bunching similar to the case of a quantum beamsplitter,
where the two polarization modes take over the role of the two input
ports. We found that the photon statistics $P_{n}$ can be calculated
best by projection on the required Fock states using polarization-rotated
Fock space ladder operators $b_{x/y}^{\dagger}=a_{x/y}^{\dagger}\cos\theta_{out}\mp a_{y/x}^{\dagger}\sin\theta_{out}$,
and tracing out the undesired polarization component afterwards. For
the numerically \cite{Johansson2013} calculated photonic density
matrix operator $\rho$ of our system, the photon number distribution
after the polarizer becomes \footnote{We calculate this inside of the cavity to clearly demonstrate the
effect; in reality, the output coupler mirror would reshape the photon
number distribution before the polarizer acts.}:

\[
P_{n}=\sum_{m=0}^{N}\frac{1}{n!\,m!}\langle0_{x}0_{y}|\left(b_{x}\right)^{n}\left(b_{y}\right)^{m}\,\rho\,\left(b_{x}^{\dagger}\right)^{n}\left(b_{y}^{\dagger}\right)^{m}|0_{x}0_{y}\rangle
\]

\begin{figure}
\includegraphics[width=1\columnwidth]{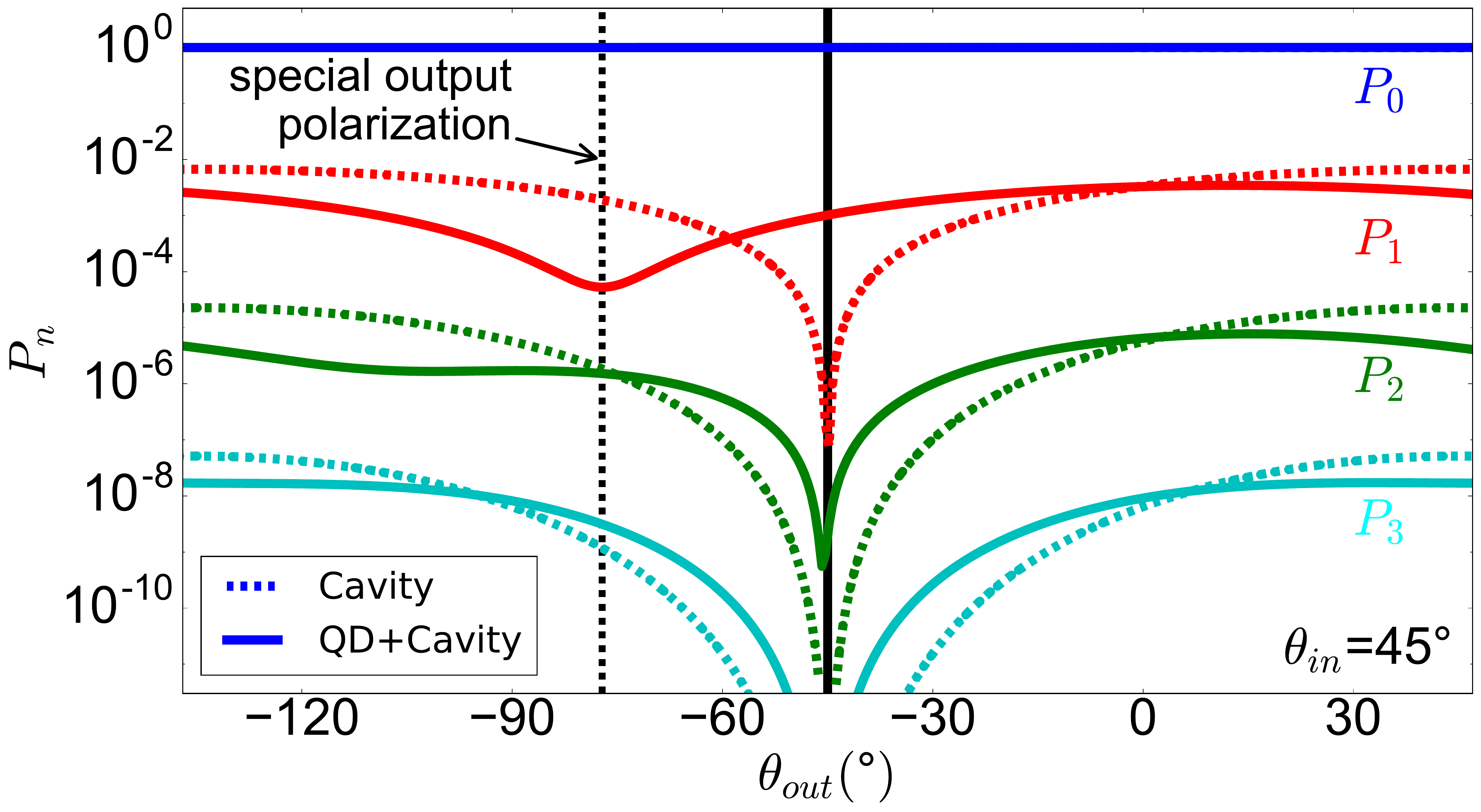}

\caption{\label{fig:Pns} Calculated photon number distribution after the polarizer,
with (through curves) and without (dashed curves) coupling to the
quantum dot in the cavity, the laser frequency is set to one of the
QD resonances. With QD, we clearly see the photon-number dependent
shift of the transmission dip. Only the photon number distribution
of the detected polarization component is shown, therefore the total
number of photons in case with quantum dot can exceed the case without
quantum dot due to polarization conversion by the dot. For clarity,
pure dephasing has been neglected here, making the special polarization
angle different from the previous simulations and experimental results.}
\end{figure}

Fig.~\ref{fig:Pns} shows the 4 lowest photon number probabilities
as a function of the polarizer angle $\theta_{out}$, for the case
with and without quantum dot. In the bare cavity case we see, as expected,
lowest transmission under the cross-polarization condition ($\theta_{out}=45^{\circ}$).
For the case with the QD, we observe a photon-number dependent shift
of the transmission dip. At the special polarization angle $\theta_{out}^{\ast}$,
we see that the one-photon component reaches a minimum while the higher-photon
number states do not. Since $g^{2}(0)\propto2\,P_{2}/P_{1}^{2}$ for
$P_{2}\ll P_{1}$ and $P_{n>2}\ll P_{2}$, the photon correlations
diverge like $g^{2}(0)\propto1/\alpha^{2}$ if the single-photon component
is attenuated as $P_{1}\rightarrow\alpha P_{1}$, explaining the enhanced
photon bunching enabled by the purification technique.

It is important to note that also the two-photon transmission dip
($P_{2}$) is not exactly at cross-polarization, which suggests the
following intuitive explanation: Apparently, in the photon number
basis, the different Fock states pick up a different phase during
transmission through the QD-cavity system. In the weak coupling regime,
but often also in the strong coupling regime, the individual Jaynes
Cummings dressed states cannot be resolved spectrally because $g\lesssim\kappa$.
However, the CQED system is still photon-number sensitive, which implies
lifetime-dependent Jaynes Cummings effects in the weak coupling regime:
the decay rate of the CQED system increases with the number of photons
in the cavity \cite{illes2015,laussy2012b}. As consequence, higher
photon-number states have a modified interaction cross section and
experience a reduced phase shift. The dip in $P_{2}$ in Fig.~\ref{fig:Pns}
is already very close to the cross-polarization angle $\theta_{out}=-45^{\circ}$,
while the dips for higher photon number states $P_{n>2}$ is indistinguishable
from $\theta_{out}=-45^{\circ}$.

We add that, while single Fock states have an undefined phase, superpositions
of different Fock states, such as in different polarization modes
have a well defined relative phase. This is also the basis of a recent
observation of quadrature squeezing in the resonance fluorescence
of a single quantum dot \cite{schulte2015}, where orthogonal polarizations
have been used to obtain a phase-stable local oscillator. We have
calculated the amount of quadrature squeezing of the transmitted light
in our system and predict slightly larger squeezing amplitudes than
those observed in \cite{schulte2015}, which is expected due to the
cavity enhancement in our system. However, photon correlations as
strong as we report here can only be achieved by coupling a QD to
a high-quality optical cavity, which has to be sufficiently polarization
degenerate to enable our purification technique: the quantum dot transition
has to couple simultaneously to both orthogonal polarizations, for
details see the supplemental information \cite{Note1}. 

\medskip{}

In conclusion, we found that the nonlinear response of a lossy cavity-quantum
dot system can be strongly enhanced by postselection of a particular
polarization state. This leads to interference between Fock states
that experienced different modifications by the QD nonlinearity, and
results in strong photon correlations of the transmitted light. As
the underlying effect, interference of the two polarizations modes
leads to high-fidelity cancellation of the single-photon transmission
for the special polarization pre- and postselection. By correlating
the results with a theoretical model, we found indications of photon-number
sensitive Jaynes-Cummings physics in the weak coupling regime of CQED.\medskip{}

\noindent

\begin{acknowledgments}
We thank G. Nienhuis for fruitful discussions. We acknowledge funding
from FOM-NWO (08QIP6-2), from NWO/OCW as part of the Frontiers of
Nanoscience program, and from the National Science Foundation (NSF)
(0901886, 0960331).
\end{acknowledgments}

\newpage{}

\onecolumngrid%
\begin{center}
\textbf{Purification of a single photon nonlinearity\\
Supplemental Material}
\end{center}

\setcounter{table}{0}
\renewcommand{\thetable}{S\arabic{table}}
\setcounter{figure}{0}
\renewcommand{\thefigure}{S\arabic{figure}}%

\section{Sample structure}

The samples under study are grown by molecular beam epitaxy on a GaAs
{[}100{]} substrate. Two distributed Bragg reflectors (DBR) surround
a $\sim5\lambda$ thick cavity \cite{Bakker2015e} containing in the
center InGaAs self-assembled quantum dots (QDs) and an oxide aperture
for transverse confinement. The top DBR mirror consists of 26 pairs
of $\lambda/4$ thick GaAs / Al$_{0.90}$Ga$_{0.10}$As layers, while
the bottom mirror has 13 pairs of GaAs / AlAs layers and 16 pairs
of GaAs / Al$_{0.90}$Ga$_{0.10}$As layers. The aperture is made
of a 10~nm thick AlAs layer which is embedded between 95 nm Al$_{0.83}$Ga$_{0.17}$As
and 66 nm thick Al$_{0.75}$Ga$_{0.25}$As. After wet chemical oxidation
this enables an intra-cavity lens for transverse mode confinement.
In the paper, we discuss two quantum dots, QD A and QD B, in two different
samples. QD A is separated by a 20 nm thick GaAs tunnel barrier from
the n-doped GaAs:Si ($2.0\times10^{18}$ cm$^{-3}$), while QD B is
separated by a 35~nm tunnel barrier to reduce co-tunneling induced
decoherence.

\section{Theoretical modelling}

\subsection{Jaynes-cumming quantum master equation}

We describe the QD-cavity system via an extended version of a two
level system in an optical cavity, which is driven by a classical
coherent laser field. The quantum description, based on the application
of a unitary transformation to transform the Hamiltonian from a time
dependent to a time independent form and the rotating wave approximation,
results in the following Hamiltonian $(\hbar=1)$\cite{Jaynes1963,Arnold2015}:

\begin{eqnarray*}
H & = & \left(\omega_{L}-\omega_{c}\right)\hat{a}_{X}^{\dagger}\hat{a}_{X}+\left(\omega_{L}-\omega_{c}\right)\hat{a}_{Y}^{\dagger}\hat{a}_{Y}+\left(\omega_{L}-\omega_{QD}^{X}\right)\hat{\sigma}_{X}^{\dagger}\hat{\sigma}_{X}+\left(\omega_{L}-\omega_{QD}^{Y}\right)\hat{\sigma}_{Y}^{\dagger}\hat{\sigma}_{Y}\\
 &  & +g_{Y}\left(\hat{\sigma}_{Y}\hat{a}_{Y}^{\dagger}+\hat{\sigma}_{Y}^{\dagger}\hat{a}_{Y}\right)+g_{X}\left(\hat{\sigma}_{X}\hat{a}_{X}^{\dagger}+\hat{\sigma}_{X}^{\dagger}\hat{a}_{X}\right)+\frac{\eta}{2}\left[e_{x}^{\prime}\left(\hat{a}_{X}^{\dagger}+\hat{a}_{X}\right)+e_{y}^{\prime}\left(\hat{a}_{Y}^{\dagger}+\hat{a}_{Y}\right)\right]
\end{eqnarray*}
Here $\omega_{c}$ is the cavity resonance frequency and $\omega_{QD}^{X/Y}$
are the fine-structure-split QD transition frequencies. $\hat{a}_{X/Y}^{\dagger}$
is the photon creation operator for a photon in X/Y polarization,
and $\hat{\sigma}_{X/Y}^{\dagger}$ creates an X/Y polarized neutral
exciton. The terms with coupling constants $g_{X/Y}$ describe the
interaction between a QD transition and the cavity field. This Hamiltonian
is designed for a polarization degenerate cavity. The last term describes
the driving of the cavity by an external linearly polarized coherent
laser field, where $\eta^{2}$ is proportional to the incident intensity
\cite{Tang2015a}, and the Jones vector $\left(e_{x}^{\prime},e_{y}^{\prime}\right)$
describes the incident light polarization.

Next we write down a quantum master equation for our Hamiltonian and
include Lindblad-type dissipation for the cavity decay rate $\kappa$,
the population relaxation rate $\gamma_{||}$ and the total pure dephasing
rate $\gamma^{\ast}$. 
\begin{equation}
\frac{d\rho}{dt}=\mathfrak{L}\rho=-i\left[\hat{H},\rho\right]+\sum_{j=X,Y}\frac{\kappa}{2}\mathfrak{D}[\hat{a}_{j}]\rho+\frac{\gamma_{||}}{2}\mathfrak{D}[\hat{\sigma_{j}}]\rho+\frac{\gamma^{*}}{4}\mathfrak{D}[\hat{\sigma}_{zj}]\rho,\label{eq:master}
\end{equation}
Where $\rho$ is the density matrix of the QD-cavity system, $\mathfrak{L}$
is the Liouvillian superopererator for QD-cavity density matrix and
$\mathfrak{D}[\hat{o}]\rho\equiv2\hat{o}\rho\hat{o}^{\dagger}{}-\hat{o}^{\dagger}{}\hat{o}\rho-\rho\hat{o}^{\dagger}{}\hat{o}$
results in Lindblad type dissipation. Here $\hat{\sigma}_{zj}$ is
defined as $\frac{1}{2}\left(\hat{\sigma}_{j}^{\dagger}\hat{\sigma}_{j}-\hat{\sigma}_{j}\hat{\sigma}_{j}^{\dagger}\right)$.
This Lindblad-type master equatition in Eq.~\ref{eq:master} is based
on the validity of several additional approximations (see for instance
\cite{Johansson2012a}), where we point out a few: (1) full separability
of the system and the environment at $t=0$, and (2) the state of
the environment does not change significantly under interaction with
the system, i.e., the interaction is weak, and the system and environment
remain separable throughout the evolution. Last, we assume (3) that
the environment has no memory on the time scale of the system (Markov
approximation). Those approximations are justified as we only discuss
photonic interaction with the environment here, which is very weak.
We are interested in the steady state solution for $\rho$, and solve
$\mathfrak{L}\rho=0$, using the numerical methods provided by the
software package QUTIP \cite{Johansson2013}.

\subsection{Transmission and photon correlations}

The cavity transmittivity is calculated by $T=\mathrm{Tr}\left[\rho_{0}\left(e_{1}\hat{a}_{X}^{\dagger}+e_{2}\hat{a}_{Y}^{\dagger}\right)\left(e_{1}\hat{a}_{X}+e_{2}\hat{a}_{Y}\right)\right]=\mathrm{Tr}\left(\rho_{0}\hat{a}^{\dagger}\hat{a}\right)$,
where $\left(e_{1},e_{2}\right)$ describes the output polarizer Jones
vector, and $\rho_{0}$ is the steady-state density matrix of the
system. We investigate the photon correlations by calculating the
second-order correlation function, which is independent of mirror
loss and can therefore be calculated directly from the intracavity
photon operators $\langle\hat{a}^{\dagger}\hat{a}\rangle$. The second-order
correlation function is given by $g^{2}(\tau)=\frac{\langle\hat{a}^{\dagger}(0)\hat{a}^{\dagger}(\tau)\hat{a}(\tau)\hat{a}(0)\rangle}{\langle\hat{a}^{\dagger}(0)\hat{a}(0)\rangle^{2}}$
with the time dependent photon creation operator $\hat{a}^{\dagger}(\tau)$.
In order to solve the time dependence of the operator $\hat{a}^{\dagger}(\tau)$,
we assume that the effect of the operator $\mathfrak{L}$ is small
and the eigenvalues are nondegenerate, which allows us to write $\hat{a}^{\dagger}(\tau)$
as $\hat{a}^{\dagger}e^{\mathfrak{L}\tau}.$ The effect of the operator
$\mathfrak{L}$ is small if it acts on a steady-state density matrix
\cite{Gardiner2004}.

\section{Estimation of model parameters}

\begin{figure}[!h]
\includegraphics[width=0.5\textwidth]{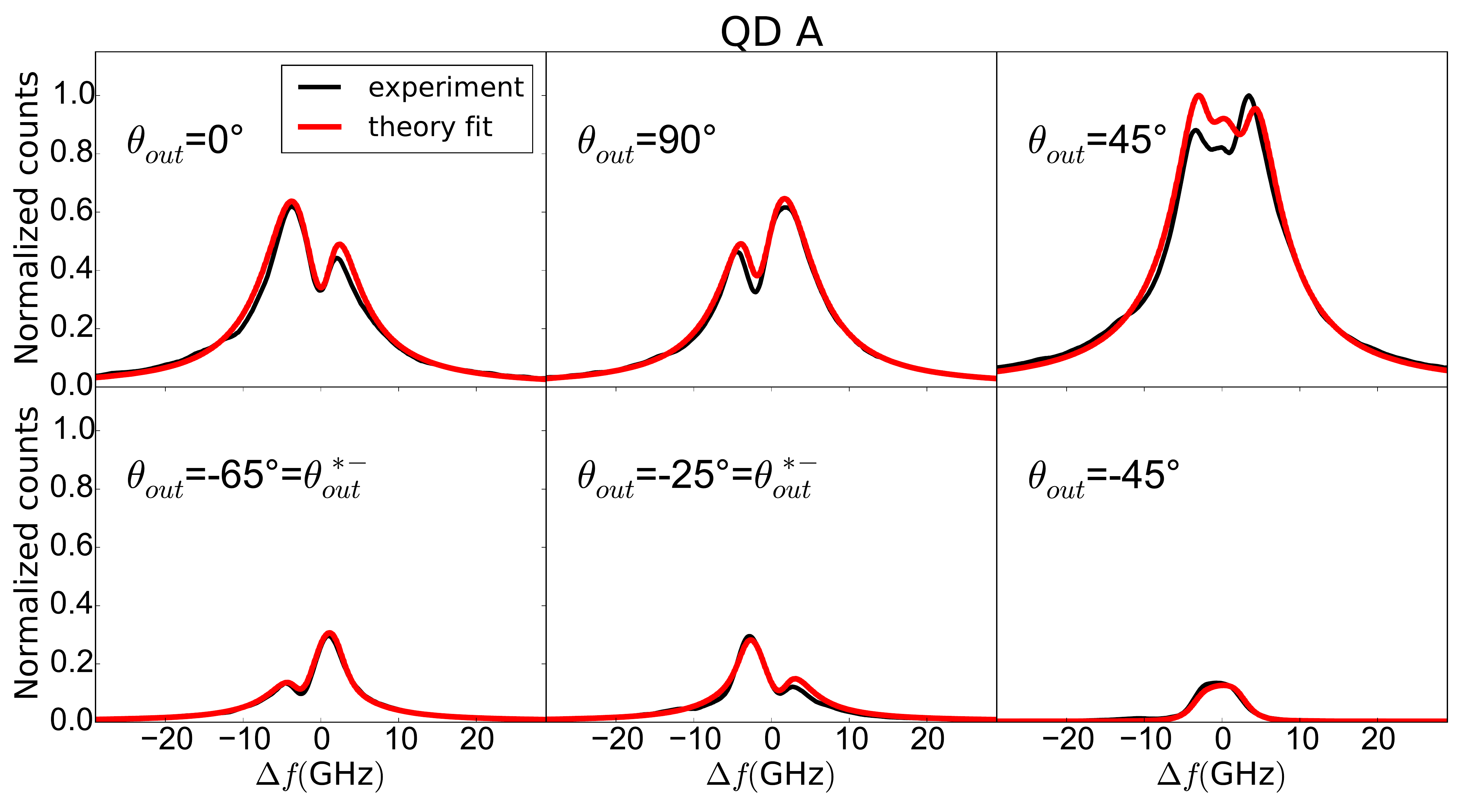}\includegraphics[width=0.5\textwidth]{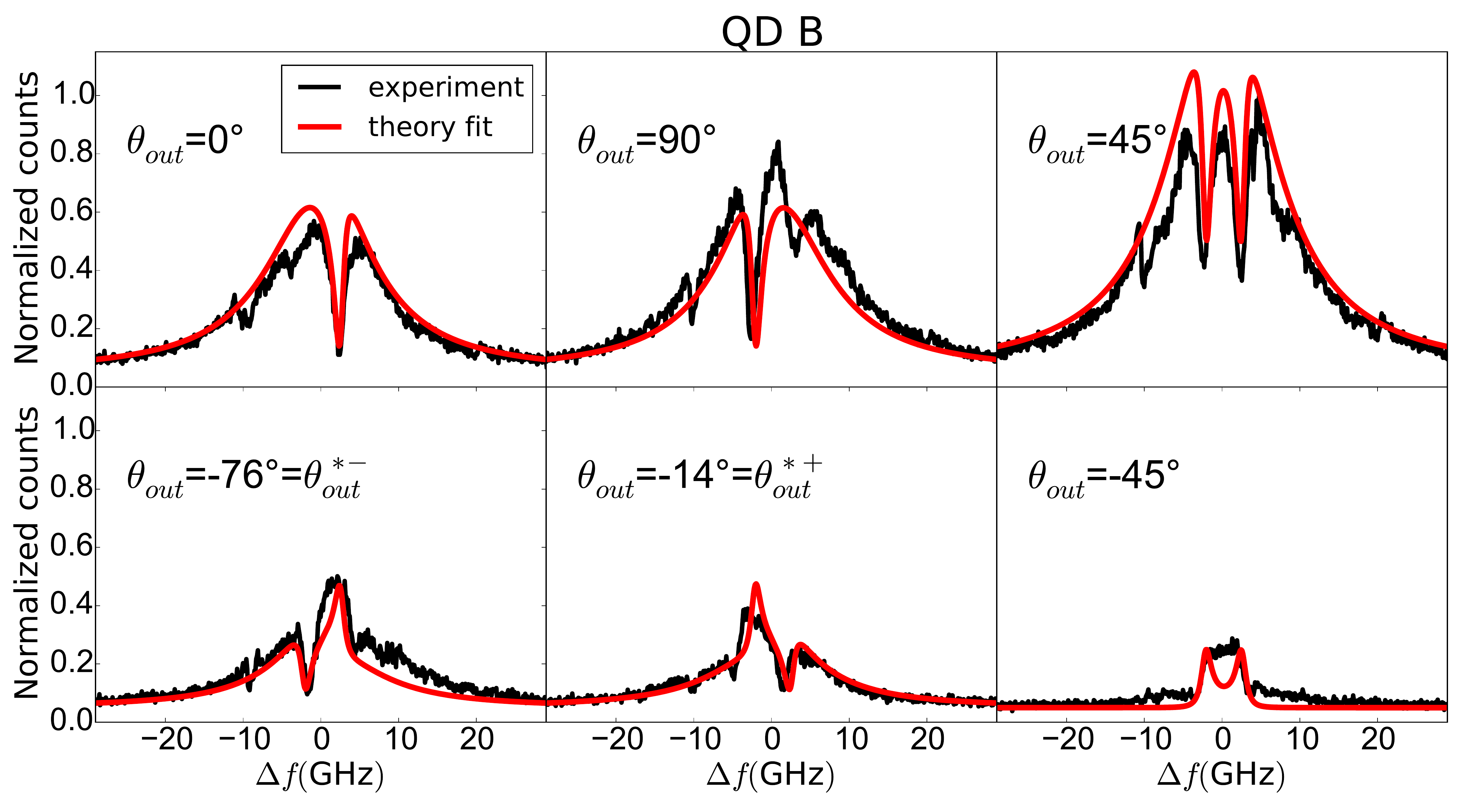}

\caption{\label{fig:fitting}Experimental data (black) and the theoretical
fit (red) for QD A (left panel) and QD B (right panel). The input
polarization was set to $\theta_{in}=45^{\circ}$ for both cases and
$\theta_{out}^{*+}$ and $\theta_{out}^{*-}$ indicate the special
polarization angles. }
\end{figure}

For estimation of the parameters, we fit the theory above discussed
to the experimental transmission data for 6 different output polarizations
for $\theta_{in}=45^{\circ}$ (excitation of both QD transitions).
The result in Fig. \ref{fig:fitting} shows excellent agreement between
experiment (blue curve) and theory (red curve). We obtain for QD A
the best-fit parameters $\kappa=69$ ns$^{-1}$, $g=15$ ns$^{-1}$,
$\gamma^{||}=3.5$ ns$^{-1}$, $\gamma^{*}=6$ ns$^{-1}$, $f_{QD}^{X/Y}=-1.5/1.3$
GHz. 

For QD B, we obtained the fitting parameters $\kappa=105$ ns$^{-1}$,
$g=14$ ns$^{-1}$, $\gamma^{||}\thickapprox1.0$ ns$^{-1}$, $\gamma^{*}\thickapprox0.7$
ns$^{-1}$, $f_{QD}^{X/Y}=-2.0/2.4$ GHz. As we discuss below, the
strongly reduced pure dephasing rates leads to much higher photon
correlations for QD B. The experimental results for QD B do not corresponds
as nice to the theory as those of QD A; the reason for this is that
the cavity of QD B is not fully polarization degenerate. This is most
clear under the cross polarization configuration ($\theta_{out}=-45^{\circ}$)
at around $\Delta f=0$. Furthermore, the fits for QD B are complicated
by the presence of another QD in the same cavity ($\theta_{out}=90^{\circ}$,
at around -10 GHz). Notice that due to different system parameters
the special polarization angles for QD A and QD B differ by $11^{\circ}.$

\newpage{}

\section{Additional photon correlation data}

\begin{figure}[h]
\includegraphics[width=6.5cm]{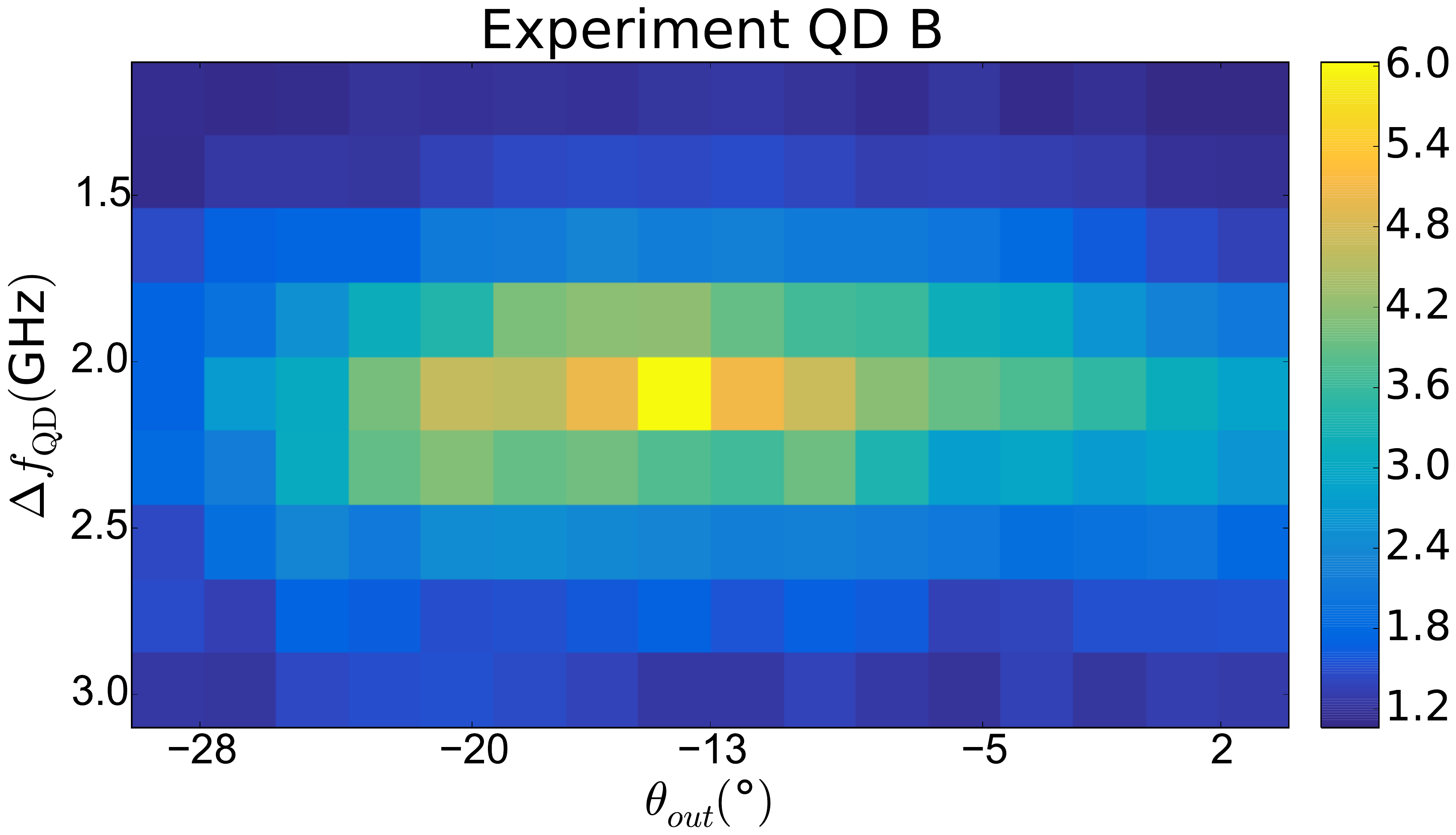}\includegraphics[width=7cm]{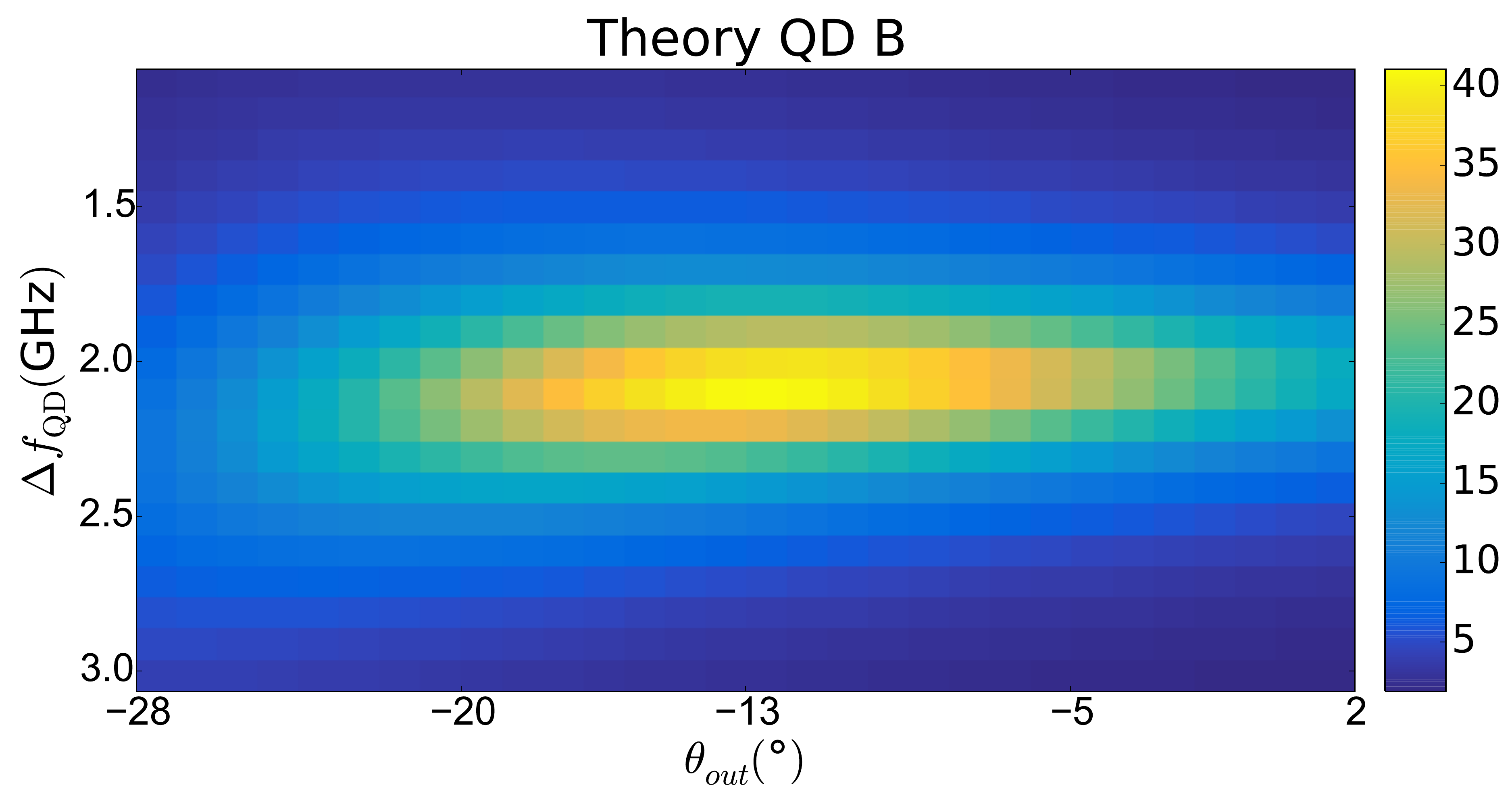}

\caption{\label{fig:qdbg22d}Experimental (left) and theoretical (right) two-photon
correlation $g^{2}(0)$ as a function of QD frequency $\Delta f_{QD}$
and output polarization $\theta_{out}$. The experimental values of
$g^{2}(0)$ are reduced by the detector jitter of 500 ps.}
\end{figure}

Here we show additional comparison of measured and calculated photon
correlation data for QD B. This data supports Fig.~1c in the main
text, and is obtained analogous to the data presented in Fig.~3.
Fig. \ref{fig:qdbg22d} shows the false-color map of $g^{2}(0)$ as
function of the output polarization $\theta_{out}$ and QD detuning.
Similar to QD A (Fig.~3) we again obtain good agreement between experiment
and theory and observe strong photon bunching for the special polarization
case. The main difference between QD A (Fig.~3) and QD B (Fig.~\ref{fig:qdbg22d})
is that, here for QD B, the obtained bunching is much larger. Again,
the difference between theory and experiment in absolute numbers in
Fig. \ref{fig:qdbg22d} is due to the detector jitter (see below Fig.
\ref{fig:convolution}).

\medskip{}

\begin{figure}[!h]
\includegraphics[width=0.8\textwidth]{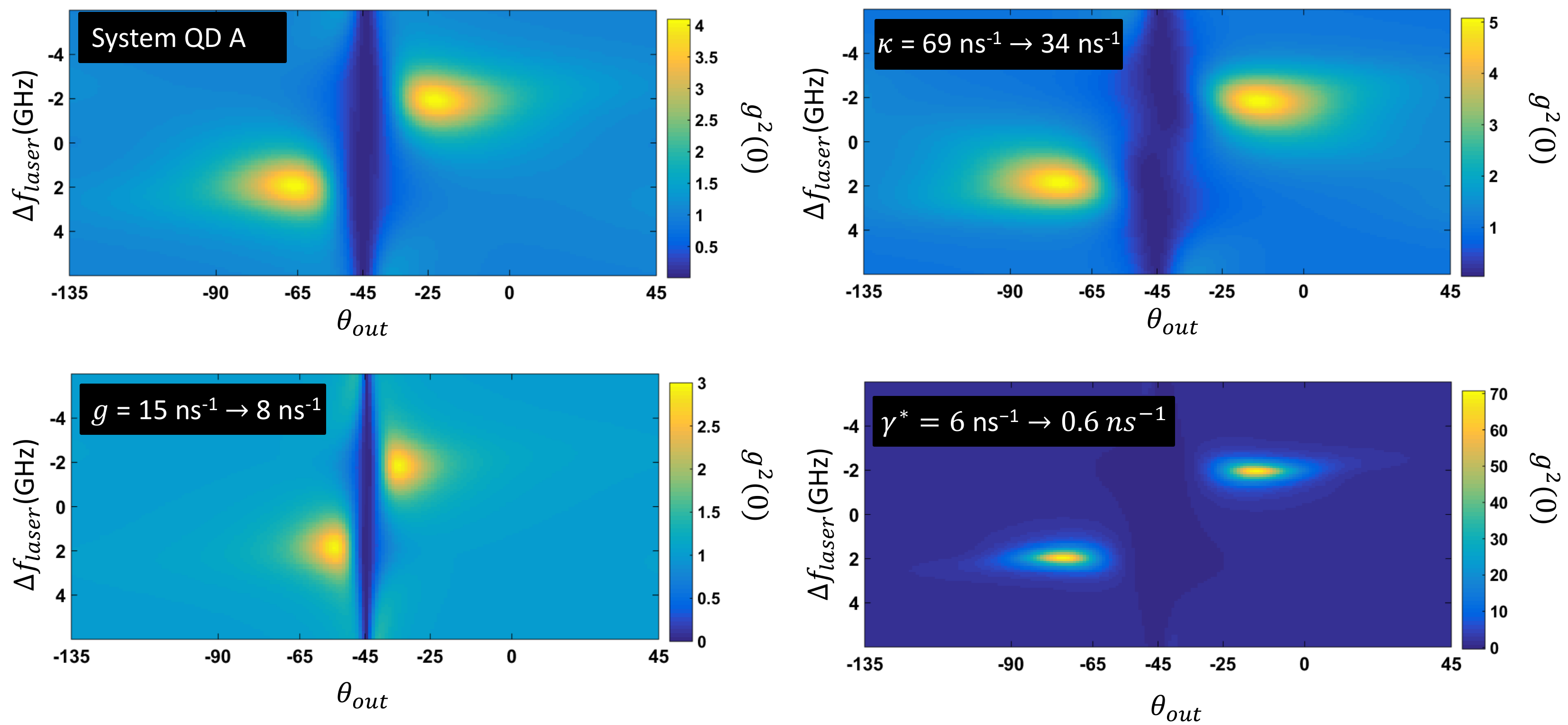}

\caption{\label{fig:qdbg2parameters}Predicted false-color plots of $g^{2}(0)$
for different system parameters. In each plot one parameter is changed
to a different value compared to the best fitting parameters for QD
A. The changed parameters are indicated in the plots. Note the changes
in the color scales.}
\end{figure}

Next, we address the question, which parameters are responsible for
the much stronger photon bunching obtained for QD B compared to QD
A? To investigate this, we have performed numerical experiments tuning
the system parameters, shown in Fig.~\ref{fig:qdbg2parameters}. 

In the false color plots of Fig. \ref{fig:qdbg2parameters} we show
the theoretical prediction of $g^{2}(0)$ of QD A while scanning the
laser over the cavity resonance and changing the output polarization
$\theta_{out}$. We show that by decreasing $\kappa$ and $g$ to
half of their original value, the effect on $g^{2}(0)$ is marginal
as compared to when we change the pure dephasing $\gamma^{*}$. A
smaller $g$ gives a slightly smaller photon bunching and a lower
$\kappa$ gives a slightly larger bunching simply because the cooperativity
increases or decreases. However, once the pure dephasing $\gamma^{*}$
is reduced from 6 to 0.5, $g^{2}(0)$ increases to a value of $\approx$70.
This corresponds well to the measurement of QD B in Fig.~1c where
we have a pure dephasing of $\gamma^{*}\thickapprox0.7$ and measure
a $g^{2}(0)\approx25$, which correponds to $g^{2}(0)\gtrsim40$ after
deconvolution. We note that for these extreme cases of photon bunching,
the exact $g^{2}(0)$ value is very sensitive to the exact model parameters
in the numeric calculations. 

We conclude that small pure dephasing is key to obtain strong photon
bunching in CQED systems.

\newpage{}

\section{QD levels}

\begin{figure}[h]
\includegraphics[scale=0.5]{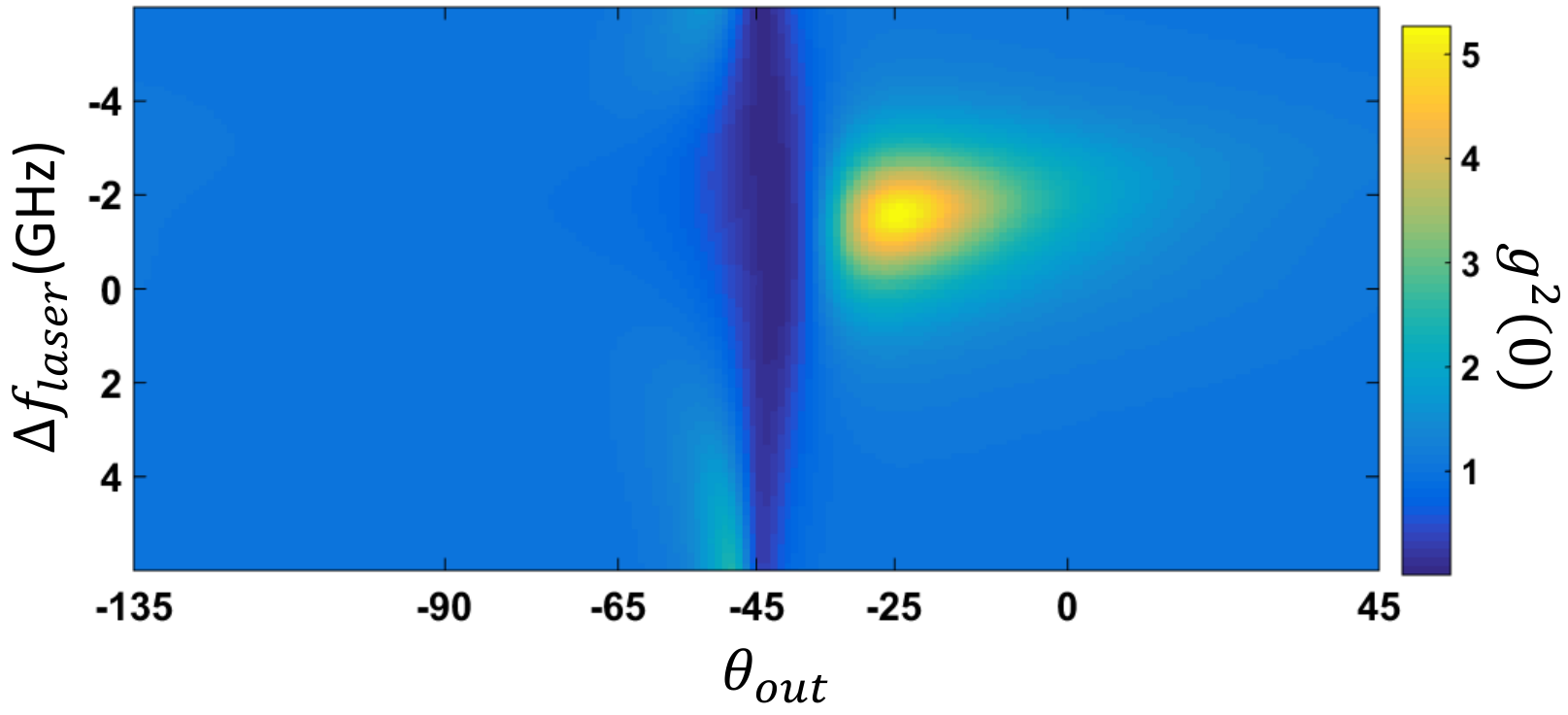}

\caption{\label{oneaxis}Theoretical data of the second-order correlation function
as a function of the QD frequency and output polarization if one of
the QD transition is removed from the simulations.}
\end{figure}

In our neutral quantum dots, the electron-hole exchange interaction
leads to two fine-structure-split optical transitions. Fig \ref{oneaxis}
shows the prediction for $g^{2}(0)$ when one QD transitions is removed.
We see, as expected, that the strong photon bunching now appears only
around the remaining QD transition, where the purification mechanism
operates. However, with a hypothetical QD without fine structure splitting,
the two QD transitions would have exactly the same energy and the
purification mechanism does not work, precluding the observation of
strong photon bunching.

\section{Detector response}

\begin{figure}[h]
\includegraphics[width=6.2cm]{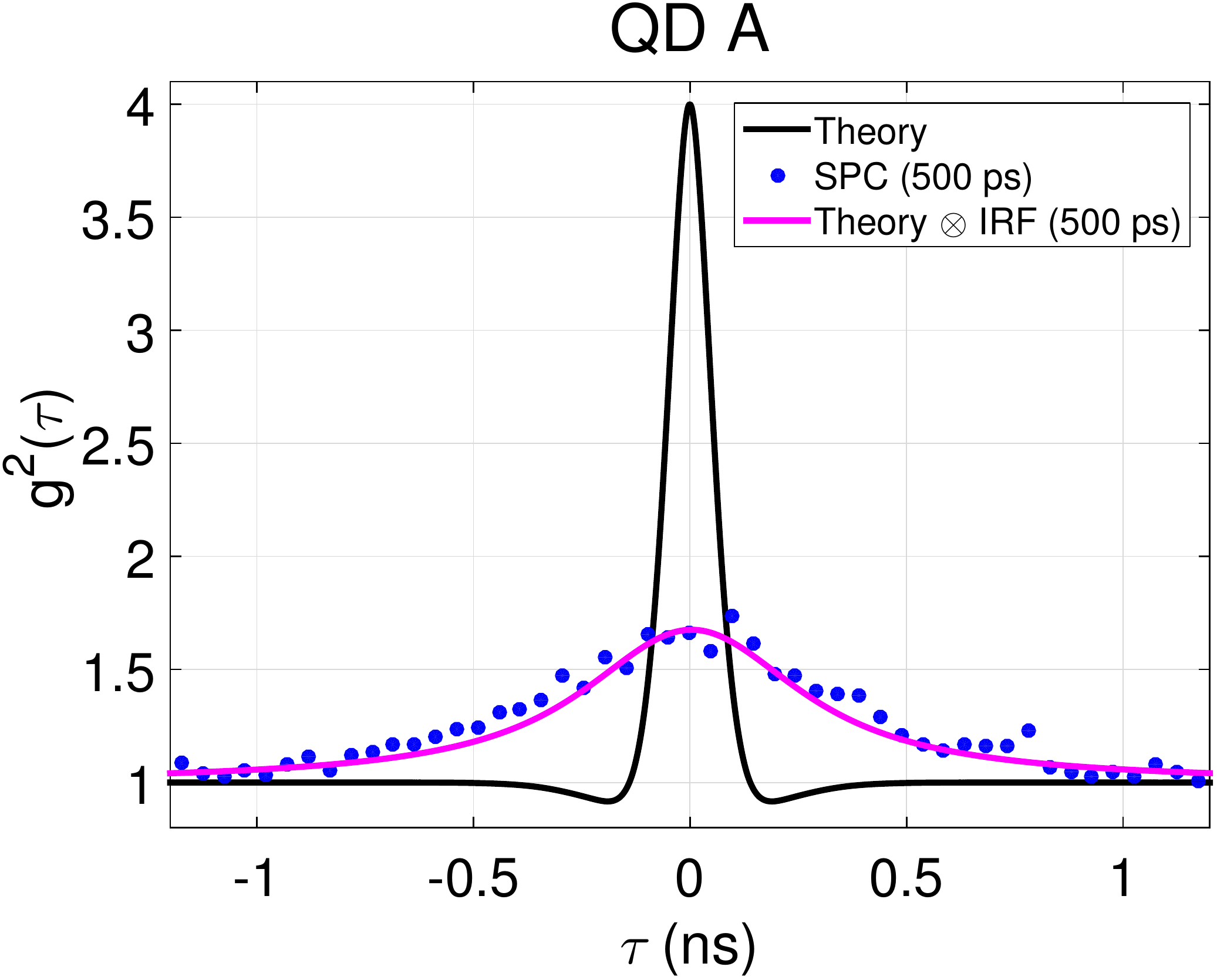}\includegraphics[width=7cm]{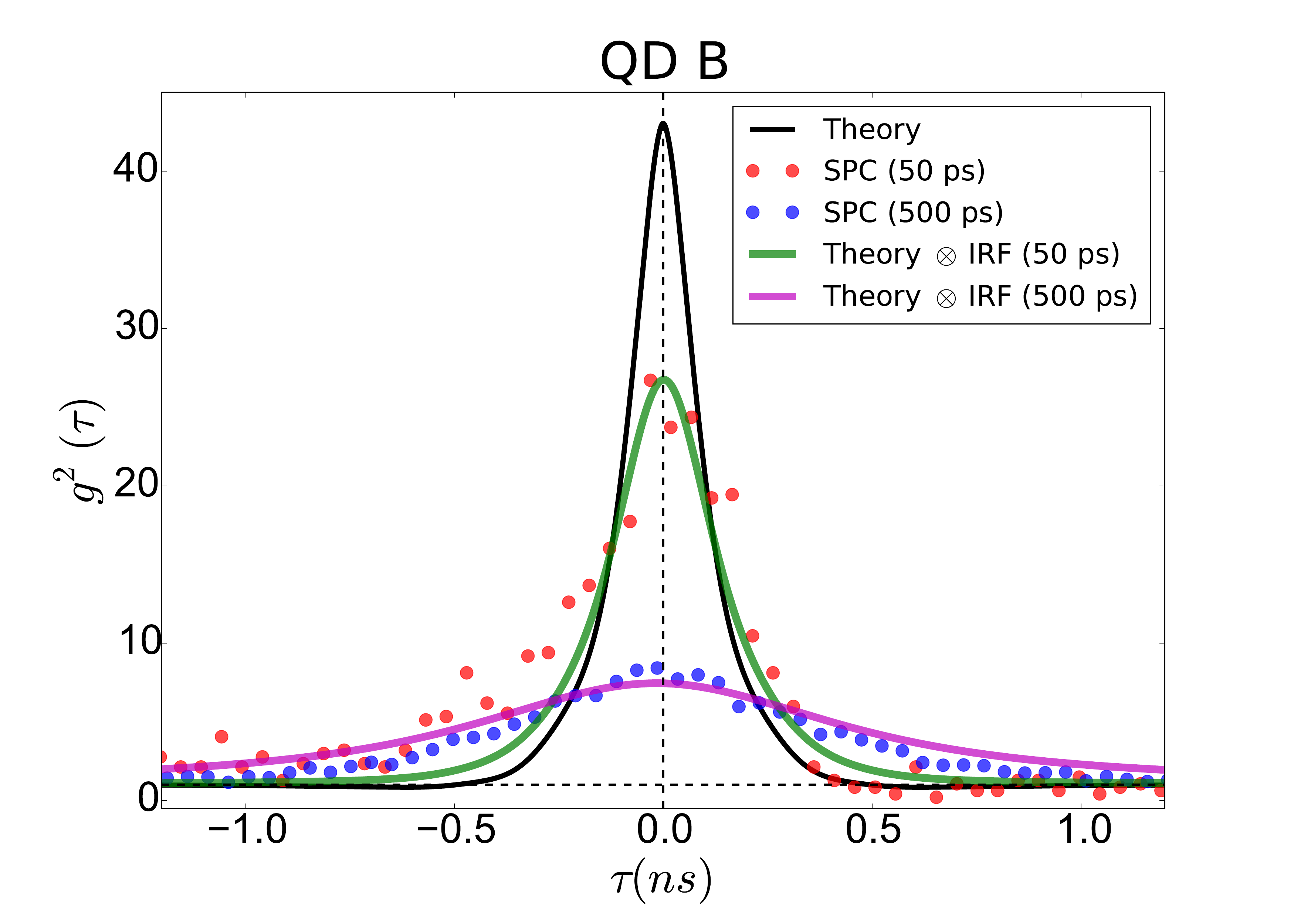}

\caption{\label{fig:convolution}Comparison of the theoretical data with and
without taking care of detector jitter, and the experimental $g^{2}(\tau)$
data for QD A (left) and QD B (right). The agreement between theory
and experiment is excellent.}
\end{figure}

In order to show that the true two-photon correlations are much stronger
than the raw experimental data suggests, we now present details on
the convolution of the theoretical $g^{2}(\tau)$ data with the single
photon counter (SPC) detector response. We use two detectors with
50~ps and 500~ps detector jitter, which was determined by measuring
photon correlations of a picosecond Ti:Sapphire laser oscillator.
As shown in Fig \ref{fig:convolution} we observe very good agreement
between the convoluted theoretical prediction and the experimental
data for both QD A en QD B. Since count rates were higher for QD B,
we could also perform the experiment with a less sensitive 50~ps
jitter detector, which again agrees very well to theory. This clearly
shows that our $g^{2}(\tau)$ measurements are severly reduced by
the detector jitter of the single photon counters, but that we can
fully deconvolute this effect.

\newpage{}

\section{Photon correlations and cavity quality}

\begin{figure}[!h]
\includegraphics[width=8cm]{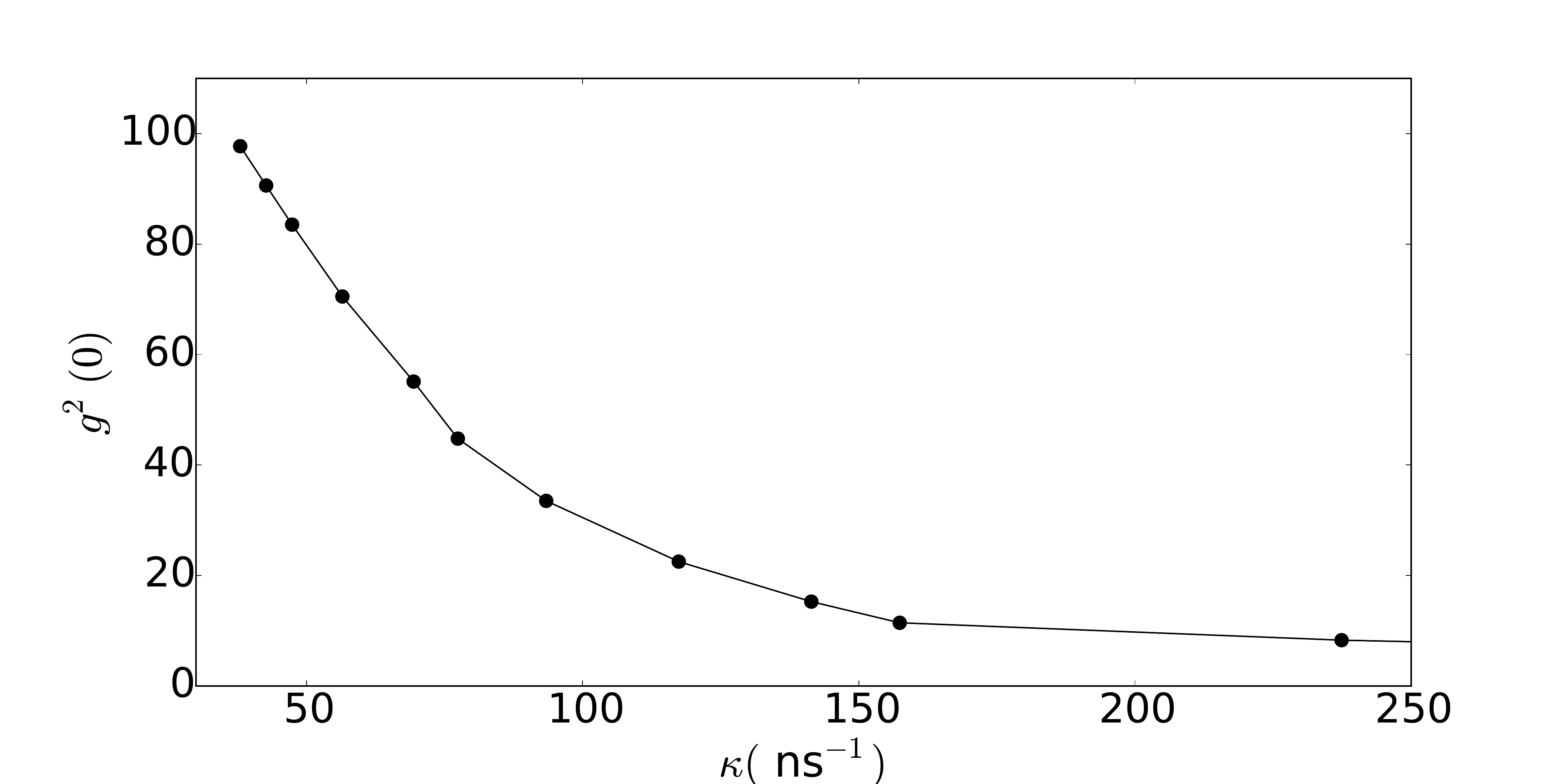}

\caption{\label{fig:kappavsg2} Calculated maximal (i.e., for special polarizer
angles) $g^{2}(0)$ for different cavity decay rates. A good cavity
with low $\kappa$ is needed in order to reach the extreme bunching
values $g^{2}(0)$. }
\end{figure}

Here we show that the cavity is essential to obtain such strong photon
correlations as we have observed experimentally. For this we conduct
numerical simulations for various cavity decay rates $\kappa$. In
order to isolate the effect of $\kappa$, we have to optimize for
each value of $\kappa$ the laser frequency and the output polarization
to find the special polarization angle and thereby the maximum in
the $g^{2}(0)$ landscape. Next to this we also need to keep the internal
mean photon number constant by increasing the incident laser power
for a higher value of $\kappa$. In order to do this we optimized
the power coupling parameter $\eta$ for each value of $\kappa$,
so that the mean photon number of the outgoing light (for parallel
polarization $\theta_{in}=\theta_{out}=45^{\circ}$) on the cavity
resonance for an empty cavity remains constant. The result is shown
in Fig. \ref{fig:kappavsg2}: In the case of almost no cavity (large
$\kappa$), only very small $g^{2}(0)$ values are obtainable, while
in good cavities (small $\kappa$), extreme values of $g^{2}(0)$
are possible. The other parameters for simulation of Fig.~\ref{fig:kappavsg2}
are similar to those of QD B.

\end{document}